\documentclass[12pt]{article}
\begin{document}
\title{\normalsize \hfill UWThPh-2000-47 \\[1cm] \LARGE
Testing CPT invariance by using C-even \\
neutral-meson--antimeson correlated states}
\author{G.\ V.\ Dass \\
\small Physics Department, Indian Institute of Technology \\
\small Powai, Bombay 400076, India \\*[3.6mm]
W.\ Grimus \\
\small Institut f\"ur Theoretische Physik, Universit\"at Wien \\
\small Boltzmanngasse 5, A--1090 Wien, Austria \\*[3.6mm]
L.\ Lavoura \\
\small Universidade T\'ecnica de Lisboa \\
\small Centro de F\'\i sica das Interac\c c\~oes Fundamentais \\
\small Instituto Superior T\'ecnico, P--1049-001 Lisboa, Portugal \\*[4.6mm] }

\date{12 December 2000}

\maketitle

\begin{abstract}
We consider the decays
of a correlated neutral-meson--antimeson state
with C-parity $+1$.
We show that there is CPT noninvariance
in the mixing of the neutral mesons if,
for any two decay modes $f$ and $g$,
the decay rate has a component $R_A$ which is antisymmetric
under the interchange of the decay times $t_1$ and $t_2$.
In particular,
one may cleanly extract the CPT-noninvariance parameter
with the help of $R_A$
by using opposite-sign dilepton events.

\vspace*{8mm}

\normalsize\noindent
PACS numbers: 11.30.Er, 13.20.-v
\end{abstract}

\newpage

The system formed by a spin-0 flavoured neutral meson $M^0$
and its antimeson $\bar M^0$
(where $M$ may be either $K$,
$D$,
$B_d$,
or $B_s$)
is experimentally interesting for testing the symmetries CP,
T, and CPT \cite{testing}.
We may recall that CP violation was first observed
in the $K^0$--$\bar K^0$ system \cite{cronin},
and is now being actively searched for
in the $B^0_d$--$\bar B^0_d$ system \cite{babar};
there are also observations of T violation \cite{CPLEAR1}
and tests of CPT invariance \cite{CPLEAR2}
in the $K^0$--$\bar K^0$ system,
although their interpretation remains controversial \cite{controversy}.

We introduce the usual propagation eigenstates
\begin{equation}
\label{MHL}
| M_H \rangle = p_H | M^0 \rangle + q_H | \bar M^0 \rangle
\quad \mbox{and} \quad
| M_L \rangle = p_L | M^0 \rangle - q_L | \bar M^0 \rangle
\, .
\end{equation}
These states get multiplied by probability amplitudes
$\exp \left( - i \lambda_H t \right)$
and $\exp \left( - i \lambda_L t \right)$,
respectively,
for nonzero proper time $t$.
Here,
\begin{equation}
\label{lambdaHL}
\lambda_H = m_H - {\textstyle \frac{i}{2}}\, \Gamma_H
\quad \mbox{and} \quad
\lambda_L = m_L - {\textstyle \frac{i}{2}}\, \Gamma_L \, ,
\end{equation}
where $m_H$,
$m_L$,
$\Gamma_H$,
and $\Gamma_L$ are real.

Defining \cite{luis}
\begin{equation}
\theta = \frac{q_H/p_H - q_L/p_L}{q_H/p_H + q_L/p_L}
\quad \mbox{and} \quad
\frac{q}{p}= \sqrt{\frac{q_H q_L}
{p_H p_L}}\, ,
\end{equation}
one finds that both the real and the imaginary parts of $\theta$
are in principle measurable and violate both CP and CPT;\footnote{For
discussions of other CPT-noninvariant observables, see for instance the recent
reviews in Ref.~\cite{review}, and the references cited therein.}
while the phase of $q/p$ is convention-dependent and,
therefore,
unphysical.
On the other hand,
the magnitude of $q/p$ is measurable and signals T and CP violation
in the mixing when it differs from 1.

It is convenient to define the functions
\begin{equation}
\label{g}
g_\pm (t) = {\textstyle \frac{1}{2}}
\left( e^{- i \lambda_H t} \pm e^{- i \lambda_L t} \right) .
\end{equation}
Then,
the probability amplitudes $a(t)$,
$b(t)$,
$\bar a(t)$,
and $\bar b(t)$ for,
respectively,
the transitions $M \rightarrow M$,
$M \rightarrow \bar M$,
$\bar M \rightarrow \bar M$,
and $\bar M \rightarrow M$ are \cite{luis,sanda}
\begin{equation}
\label{ab}
\begin{array}{lll}
a(t) & = & g_+(t) - \theta g_-(t) \, , \\*[2mm]
b(t) & = & {\displaystyle \frac{q}{p}}\, \sqrt{1-\theta^2} g_-(t) \, , \\*[2mm]
\bar a(t) & = & g_+(t) + \theta g_-(t) \, , \\*[2mm]
\bar b(t) & = & {\displaystyle \frac{p}{q}}\, \sqrt{1-\theta^2} g_-(t) \, .
\end{array}
\end{equation}

In the experimental study of the neutral-meson--antimeson systems
it is interesting to use
correlated meson--antimeson states of the form
\begin{equation}
\label{+}
| M_\pm \rangle = {\textstyle \frac{1}{\sqrt{2}}}
\left[
| M^0 ( \vec k ) \rangle \otimes | \bar M^0 ( - \vec k ) \rangle \pm
| \bar M^0 ( \vec k ) \rangle \otimes | M^0 ( - \vec k ) \rangle
\right] ,
\end{equation}
in which one of the mesons flies in one direction
(denoted by the momentum $\vec k$)
and the other one flies in the opposite direction.
Indeed,
the DA$\Phi$NE experiment at Frascati will use correlated states
of the form $| K_- \rangle$,
while the Belle and BaBar Collaborations are working
with states $| B_{d-} \rangle$.
The correlated states $| M_\pm \rangle$ have C-parity $\pm 1$
and are produced from the decay of certain spin-1 resonances
like the $\Phi$ and $\Upsilon (4S)$.
At $e^+ e^-$ colliders, states of the type
$| M_- \rangle$
are preferred since the produced resonances almost always have
quantum numbers $J^{PC} = 1^{--}$.
Possible ways of experimentally producing states
of the type $| M_+ \rangle$
have nevertheless been discussed in the literature,
see for instance Ref.~\cite{ways}.

It has recently been shown \cite{luis} that
it is difficult or even impossible to extract $\theta$
from experiments with states
of the experimentally preferred form $| M_- \rangle$ alone.
This happens because,
in the semileptonic decays of $| M_- \rangle$,
the effects of CPT noninvariance
are entangled with those of violations of the $\Delta F = \Delta Q$ 
rule---where
$F$ means flavour, i.e.,
$F$ may be either $S$,
$C$,
or $B$.
Thus,
it is difficult to know
whether one is really measuring
violations of CPT invariance
or one is measuring violations of that rule. For instance, in
Ref.~\cite{dass} CPT invariance has been used for checking
the rule $\Delta B = \Delta Q$, 
whereas in Ref.~\cite{OPAL}
$\theta$ has been constrained by assuming the validity of this rule.

The purpose of this note is to show that
$\theta$ can neatly be determined by using
the opposite-sign dilepton decays
of the symmetrical correlated state
$| M_+ \rangle$, which has C-parity $+1$.
As a matter of fact,
tests of CPT invariance using states $| M_+ \rangle$
do not necessarily require the use of any specific final states---like
the semileptonic decays---and the signal for CPT noninvariance
is independent of assumptions
about the decay amplitudes to the particular final states
that one may wish to use.
However,
if one wants to extract the actual value
of the CPT-violating parameter $\theta$,
then it is better to use opposite-sign semileptonic decays.

Let us consider the event in which the meson
of $| M_+ \rangle$ with momentum $\vec k$
decays into $f$ at time $t_1$,
while the meson with momentum $- \vec k$
decays into $g$ at time $t_2$.
The rate of such an event is
\begin{equation}
\label{extra formula for R}
R \left( f, t_1; g, t_2 \right) = {\textstyle \frac{1}{2}}
\left| {\cal A} \left(f, t_1; g, t_2 \right)
\right|^2 ,
\end{equation}
where
\begin{eqnarray}
{\cal A} \left(f, t_1; g, t_2 \right) &=&
\left[ a \left( t_1 \right) A_f + b \left( t_1 \right) \bar A_f \right]
\left[ \bar b \left( t_2 \right) A_g
+ \bar a \left( t_2 \right) \bar A_g) \right] \nonumber \\
& & +  \left[ \bar b \left( t_1 \right) A_f
+ \bar a \left( t_1 \right) \bar A_f \right]
\left[ a \left( t_2 \right)
A_g + b \left( t_2 \right) \bar A_g
\right] \, ,
\label{amplitude}
\end{eqnarray}
where $A_f$ is the amplitude for the decay
$| M^0 \rangle \rightarrow | f \rangle$,
$\bar A _f$ is the amplitude for
$| \bar M^0 \rangle \rightarrow | f \rangle$,
and similarly for the decays into $g$.
One may use Eqs.~(\ref{ab}),
together with
\begin{equation}
\label{ggg}
\begin{array}{lll}
g_+ \left( t_1 \right) g_+ \left( t_2 \right) +
g_- \left( t_1 \right) g_- \left( t_2 \right) &=&
g_+ \left( t_1 + t_2 \right) ,
\\*[1mm]
g_+ \left( t_1 \right) g_- \left( t_2 \right) +
g_- \left( t_1 \right) g_+ \left( t_2 \right) &=&
g_- \left( t_1 + t_2 \right) ,
\end{array}
\end{equation}
to show that the amplitude $\mathcal{A}$
given in Eq.~(\ref{amplitude}) can be written as
\begin{eqnarray}
{\cal A} \left(f, t_1; g, t_2 \right) &=&
\frac{p}{q}\, \sqrt{1 - \theta^2}
\left[ g_- \left( t_1 + t_2 \right)
- 2 \theta g_- \left( t_1 \right) g_- \left( t_2 \right)
\right] A_f A_g \nonumber \\
& & + \frac{q}{p}\, \sqrt{1 - \theta^2}
\left[ g_- \left( t_1 + t_2 \right)
+ 2 \theta g_- \left( t_1 \right) g_- \left( t_2 \right)
\right] \bar A_f \bar A_g \nonumber \\
& & + \left[ g_+ \left( t_1 + t_2 \right)
- 2 \theta^2 g_- \left( t_1 \right) g_- \left( t_2 \right)
\right] \left( A_f \bar A_g + \bar A_f A_g \right) \nonumber \\
& & + \theta \left[ g_+ \left( t_1 \right) g_- \left( t_2 \right)
- g_- \left( t_1 \right) g_+ \left( t_2 \right) \right]
\left( A_f \bar A_g - \bar A_f A_g \right) .
\label{A}
\end{eqnarray}
Notice that,
when $\theta = 0$,
the amplitude ${\cal A}$ is
not only symmetric under the interchange $t_1 \leftrightarrow t_2$ but,
as a matter of fact,
it is a function only of the sum $t_1 + t_2$ \cite{sanda,xing96}.
If, however, the CPT-noninvariance parameter $\theta$ is nonzero, then
${\cal A}$ is not any more a function only
of $t_1 + t_2$ and, indeed, it acquires
a term---the one in the last line of Eq.~(\ref{A})---{\em antisymmetric}
under $t_1 \leftrightarrow t_2$.
Then,
the decay rate can be written as the sum of
a symmetric component $R_S \left( f, t_1; g, t_2 \right)$
and an antisymmetric component $R_A \left( f, t_1; g, t_2 \right)$,
\begin{equation}
\label{RS}
\begin{array}{lll}
R_S \left( f, t_1; g, t_2 \right) &=& {\textstyle \frac{1}{2}}
\left[ R \left( f, t_1; g, t_2 \right)
+ R \left( f, t_2; g, t_1 \right) \right] ,
\\*[1mm]
R_A \left( f, t_1; g, t_2 \right) &=& {\textstyle \frac{1}{2}}
\left[ R \left( f, t_1; g, t_2 \right)
- R \left( f, t_2; g, t_1 \right)
\right] \,,
\end{array}
\end{equation}
and
{\em a nonzero $R_A$ arises
only if CPT invariance does not hold for neutral-meson mixing}
treated in the usual fashion.
We conclude that,
if experimentally one finds $R_A \neq 0$,
then this signals the presence of CPT violation.
If,
on the other hand,
$R_A$ is not significantly different from zero,
then this may allow to put a bound on $\theta$.

We want to stress that a measurable CPT-violating asymmetry exists
{\em for any two different final states
$f \neq g$} provided
$A_f \bar A_g - \bar A_f A_g \neq 0$; no other assumptions
about the amplitudes $A_f$, $A_g$, $\bar A_f$, and $\bar A_g$
are needed---in particular,
about their behaviour under the transformations CP,
CPT,
and T.

As an illustration,
let us consider the particular case
of opposite-sign dilepton events \cite{xing99},
i.e.,
inclusive semileptonic decays with $f = X \ell^+ \nu_\ell$
and $g = \bar X \ell^- \bar\nu_\ell$.
We shall use the simple notation $\ell^+$ for $f$ and $\ell^-$ for $g$.
We want to show that, in this particular case,
it is possible, at least in principle,
to explicitly derive the value of $\theta$
from the observation of the time dependence of $R_A$.
Allowing for transitions which violate the $\Delta F = \Delta Q$ rule,
we introduce the rephasing-invariant quantities \cite{luis}
\begin{equation}
\label{lambdas}
\lambda_+ = \frac{q}{p}\, \frac{\bar A_+}{A_+}
\quad \mbox{and} \quad
\bar\lambda_- = \frac{p}{q}\, \frac{A_-}{\bar A_-} \, ,
\end{equation}
where $A_+ \equiv A_{\ell^+}$,
$\bar A_- \equiv \bar A_{\ell^-}$,
and so on.
As usual,
we assume that the quantitities $\theta$,
$\lambda_+$,
and $\bar \lambda_-$,
which describe `unexpected' physics,
are small,
and we confine ourselves to the first order in these quantities.
If we do this, then the amplitude ${\cal A}$ is given by
\begin{equation}
{\cal A} = A_+ \bar A_- \left\{ g_+ \left( t_1 + t_2 \right) +
\left( \lambda_+ + \bar \lambda_- \right) g_- \left( t_1 + t_2 \right)
+ \theta \left[ g_+ \left( t_1 \right) g_- \left( t_2 \right)
- g_- \left( t_1 \right) g_+ \left( t_2 \right) \right] \right\} .
\end{equation}
With the usual definitions $\Delta m = m_H - m_L$,
$\Delta \Gamma = \Gamma_H - \Gamma_L$,
and $\Gamma = \left( \Gamma_H + \Gamma_L \right) / 2$,
we obtain for the decay rate the results
\begin{eqnarray}
\label{RSt}
R_S \left( \ell^+, t_1; \ell^-, t_2 \right) &=&
{\textstyle \frac{1}{8}} \left| A_+ \right|^2 \left| \bar A_- \right|^2
\left[ e^{- \Gamma_H t_+} + e^{- \Gamma_L t_+} +
2\, e^{- \Gamma t_+} \cos \left( \Delta m t_+ \right) \right.
\nonumber\\
& & + 2 \left( e^{- \Gamma_H t_+} - e^{- \Gamma_L t_+} \right)
\mbox{Re}\, \left( \lambda_+ + \bar \lambda_- \right)
\nonumber\\
& & + \left.
4\, e^{- \Gamma t_+}\,
\sin \left( \Delta m t_+ \right)
\mbox{Im}\, \left( \lambda_+ + \bar \lambda_- \right) \right]
\end{eqnarray}
and
\begin{eqnarray}
\label{RAt}
R_A \left( \ell^+, t_1; \ell^-, t_2 \right) &=&
{\textstyle \frac{1}{2}} \left| A_+ \right|^2 \left| \bar A_- \right|^2
e^{- \Gamma t_+}
\nonumber \\
& & \times \left\{ \mbox{Re}\, \theta \left[
\sinh \left( \frac{\Delta \Gamma t_1}{2} \right)
\cos \left( \Delta m t_2 \right)
-
\cos \left( \Delta m t_1 \right)
\sinh \left( \frac{\Delta \Gamma t_2}{2} \right) \right] \right.
\nonumber  \\
& & + \left. \mbox{Im}\, \theta \left[
\cosh \left( \frac{\Delta \Gamma t_1}{2} \right)
\sin \left( \Delta m t_2 \right)
-
\sin \left( \Delta m t_1 \right)
\cosh \left( \frac{\Delta \Gamma t_2}{2} \right) \right] \right\} ,
\nonumber \\
\end{eqnarray}
where $t_+ = t_1 + t_2$.
One may note that
violations of the $\Delta F = \Delta Q$ rule appear only in $R_S$,
while the CPT-noninvariance parameter $\theta$ appears only in $R_A$,
when we take into account `small' physics to first order only.

A glance at Eq.~(\ref{RAt}) confirms that
both the real and the imaginary part of
$\theta$ can be determined using this method.
However, in practice that
determination depends on the values of $\Delta m$ and $\Delta \Gamma$.
If we use $M = B_d$,
we must set $\Delta \Gamma \approx 0$ in Eq.~(\ref{RAt}).
Then, only ${\rm Im}\, \theta$ is measurable
and the time dependence
is $\exp \left[ - \Gamma \left( t_1 + t_2 \right) \right]
\left[ \sin \left( \Delta m t_1 \right)
- \sin \left( \Delta m t_2 \right) \right]$.
On the other hand,
for $M = K$,\footnote{The usual long-lived neutral kaon $K_L$
is our heavier state $M_H$,
and the usual short-lived neutral kaon
$K_S$ corresponds to the lighter state $M_L$.}
where $\Gamma_H \ll \Gamma_L$,
we may choose $t_1$ and $t_2$ such that
$t_1 \sim 1 / \Gamma_L \ll t_2$.
Equation~(\ref{RAt}) then yields
\begin{equation}
R_A \left( \ell^+, t_1; \ell^-, t_2 \right) \simeq
{\textstyle \frac{1}{4}} \left| A_+ \right|^2 \left| \bar A_- \right|^2
e^{- \Gamma t_1 - \Gamma_H t_2}
\left[ \mbox{Re}\, \theta \cos \left( \Delta m t_1 \right) -
        \mbox{Im}\, \theta \sin \left( \Delta m t_1 \right) \right] ,
\end{equation}
which is practically independent of $t_2$
as long as $t_2 \ll 1 / \Gamma_H$.

In conclusion,
in this note we have discussed the possibility of using
the states $| M_+ \rangle$ for probing CPT invariance
in the mixing of neutral mesons.
We have shown that this is feasible for
$A_f \bar A_g -\bar A_f A_g \neq 0$, because then
the decay rate $R \left( f, t_1; g, t_2 \right)$ has a component
which is antisymmetric with respect
to the interchange $t_1 \leftrightarrow t_2$
if and only if CPT invariance does not hold.
This conclusion
is independent of the final states $f$ and $g$,
or any assumptions thereon.
We have, in particular,
considered the case of opposite-sign dilepton events;
then, in the decay rate the effects of CPT violation appear
separated from the effects of violation
of the $\Delta F = \Delta Q$ rule.
It is then possible to cleanly extract the CPT-noninvariance parameter
$\theta$,
at least when the values of $\Delta m$ and
$\Delta \Gamma$ are not too unfavourable.
We believe that,
although experimentally it might be hard
to work with states $| M_+ \rangle$,
the cleanness with which they probe CPT invariance
might make the effort worthwhile.

\section*{Acknowledgement}

L.L.\ thanks Jo\~ao P.\ Silva for an enlightening discussion on
the production of the states $| M_+ \rangle$, and for reading the manuscript.

\end{document}